# Polariton surface solitons under resonant pump

YAROSLAV V. KARTASHOV[1,2,*] AND VICTOR A. VYSLOUKH[3]

[1]*ICFO-Institut de Ciencies Fotoniques, The Barcelona Institute of Science and Technology, 08860 Castelldefels (Barcelona), Spain*
[2]*Institute of Spectroscopy, Russian Academy of Sciences, Troitsk, Moscow, 108840, Russia*
[3]*Universidad de las Américas Puebla, Santa Catarina Mártir, 72820 Puebla, México*

**We address formation of stable dissipative surface solitons in the exciton-polariton condensate in one-dimensional array of microcavity pillars under the action of localized resonant pump acting in the edge resonator. We show that localization degree and peak amplitudes of surface solitons can be effectively controlled by the pump frequency and that the allowed energy gap of periodic structure determines the energy range, where surface solitons can form. One observes bistability at sufficiently large pump amplitudes and nonlinearity-induced shift of the position of resonance peak from the allowed energy band of the periodic array into its forbidden energy gap. Growth of the spatial period of the array reduces coupling between pillars and currents from surface pillar into bulk pillars that leads to the increase of the surface soliton amplitude. Strong expansion into the depth of array occurs for pump frequencies corresponding to the middle of the allowed energy band. Surface solitons can be excited from the broadband Gaussian noise. Above certain threshold noise level solitons from stable upper branch of the bistability curve are excited, while below threshold solitons from the lower branch form.**

From theoretical point of view, linear and nonlinear surface waves forming at the interface of two materials are very intriguing objects because their properties depend in the nontrivial way on the parameters and symmetries of different uniform or micro-structured materials placed in contact, sometimes offering wide prospects for the control of shapes, stability, and localization of surface excitations. From practical point of view, surface waves are suitable for a wide range of surface characterization, sensing, and spectroscopic applications due to their evanescent wings. Especially nontrivial task is the excitation of nonlinear surface waves that was successfully accomplished in conservative optical settings after prediction that such waves can form at the interfaces of uniform and structured materials (see reviews [1-3] and references therein). The characteristic feature of the nonlinear surface wave in conservative settings is the presence of power threshold above which such wave can form and, in the case of interface with periodic material, strong dependence of wave localization on the position of its propagation constant within forbidden gap of periodic medium. Dissipative variants of the nonlinear surface waves are known too [4]. In particular, spatially localized gain in optical systems allows excitation of solitons localized on the domain with gain (see [5] for a review of this topic), even when it is provided at the interface of periodic media [6]. Unique opportunities for lasing in surface states have been also reported in truncated periodic parity-time symmetric systems [7].

Engineered potential landscapes allowing exploration of surface wave related phenomena at the interface of periodic media can be created not only in optical, but also in optoelectronic systems, such as structured microcavities with embedded quantum wells, where strong light-matter coupling leads to the formation of the exciton-polariton quasiparticles and their condensation. Developed technologies of microcavity structuring allow creation of one- and two-dimensional potential landscapes with various interfaces between periodic and uniform cavity regions [8,9]. Such potentials have already been used for demonstration of linear nontopological [10,11] and topological [12-16] polariton edge states. Polariton condensates are characterized by strong nonlinear interactions allowing to study formation of gap solitons [17-20] and nonlinear edge states [21-25]. At the same time, polariton condensates are essentially dissipative, they require external pump for their existence [9]. Thus, they offer ideal testbed for exploration of dissipative surface states. Very recently external pump was used to achieve polariton lasing in topological edge states [26].

The aim of this Letter is to show that dissipative surface solitons can be excited in the exciton-polariton condensate in one-dimensional array of microcavity pillars under the action of *resonant* pump localized in the edge resonator. In complete contrast to nonresonant pumping schemes [6,27], resonant pump allows to study bistability of surface waves and offers remarkable control over spatial localization of the excited surface waves via tuning pump frequency.

We consider the evolution of exciton-polariton condensate described by the wavefunction $\Psi(x,t)$ in the one-dimensional truncated periodic potential $\mathcal{R}(x)$ under the influence of resonant localized pump $\mathcal{H}(x)$. The evolution of the wavefunction $\Psi$ is governed by the normalized Gross-Pitaevskii equation [9]:

$$i\frac{\partial \Psi}{\partial t} = -\frac{1}{2}\frac{\partial^2 \Psi}{\partial x^2} + \mathcal{R}(x)\Psi + |\Psi|^2\Psi - i\gamma\Psi + \mathcal{H}(x)e^{i\varepsilon t}. \qquad (1)$$

Here the transverse coordinate $x$ is scaled to the characteristic length $L$; evolution time $t$ is normalized to $\hbar\varepsilon_0^{-1}$, where $\varepsilon_0 = \hbar^2/mL^2$ is the characteristic energy scale and $m$ is the effective mass; $\varepsilon$ is the pump frequency detuning from the bottom of the lower polariton dispersion branch, normalized to $\varepsilon_0\hbar^{-1}$; the depth of potential $\mathcal{R}(x)$ is scaled to the characteristic energy $\varepsilon_0$; $\gamma = \hbar/\tau\varepsilon_0$ is the strength of losses inversely proportional to the polariton lifetime $\tau$. We assume that condensate is confined in the one-dimensional array of microcavity pillars or in the modulated microcavity wire. The potential energy landscape created by the array can be modeled by the sequence of the potential wells of the width $w_0$ with adjacent wells separation $d$:

$$\mathcal{R}(x) = -p\sum_{k=0}^{N-1} \exp[-(x-kd)^6/w_0^6], \qquad (2)$$

where $N$ is the total number of pillars, $p$ is the dimensionless depth of the potential wells. Nonlinear term in Eq. (1) accounts for repulsion between polaritons, while last term takes into account spatially localized quasi-resonant pump $\mathcal{H}(x)$. In what follows, we focus our attention on narrow excitation of the left-edge microcavity pillar described by the function $\mathcal{H}(x) = he^{-x^2/w_m^2}$ with amplitude $h$ and width $w_m$ approximating the width of the fundamental mode of pillar. For characteristic length $L = 2\,\mu$m and effective mass $m = 10^{-34}$ kg one obtains characteristic energy scale of $\varepsilon_0 = 0.17$ meV and time scale $\hbar\varepsilon_0^{-1} \sim 3.8$ ps. In simulations we used experimentally relevant parameters $p = 8$, $w_0 = 0.5$, $\gamma = 0.04$, $w_m = 0.6$, $N = 31$, and concentrated on near the resonance excitation of polariton surface states under the action of resonant pump of variable amplitude $h$ for different spatial periods $d$ of the array.

First, we consider energy spectrum of the linear system, since its structure is crucially important for understanding of system's response to resonant excitation. Using Eq. (1) in the case of single microcavity pillar (or single potential well) in the absence of nonlinearity, one can estimate the amplitude $a$ and phase $\phi$ of the steady-state oscillations, $\Psi = ae^{i\phi + i\varepsilon t}$, emerging in the presence of localized pump, as $a = h(\gamma^2 + \delta^2)^{-1/2}$ and $\phi = -\arctan(\delta/\gamma)$, where $\delta = \varepsilon - \varepsilon_k$ is the pump frequency detuning from the "resonant" frequency $\varepsilon_k$ of the nearest eigenmode of the system. The pump profile is supposed to coincide with that of the corresponding linear mode. Naturally, in the array the spectrum substantially enriches in comparison with single-well case due to coupling between neighboring pillars. Such linear spectrum can be found by setting $h, \gamma = 0$ in Eq. (1) and representing solution in the form $\Psi = \psi(x)e^{i\varepsilon t}$ that leads to linear eigenvalue problem from which mode shapes and eigenvalues $\varepsilon = \varepsilon_k$ can be found. In the case of finite array with $N$ pillars the eigenvalues $\varepsilon_k$ form two bands that are separated by the broad forbidden gap. Notice that the single pillar supports two modes for our parameters. Figure 1(a) shows eigenvalues from the upper band for different spatial periods $d$. This band quickly narrows down with increase of the period $d$ due to reduction of the coupling strength between pillars [Fig. 1(b)]. Like in any externally driven system, the projection of the pump on eigenmodes of the system determines their excitation efficiency. Figure 1(c) illustrates projections $I_k = \int_{-\infty}^{+\infty} \psi_k^*(x)\mathcal{H}(x)dx$ of localized pump on different eigenmodes of loss-free system with eigenvalues $\varepsilon_k$, where $\psi_k$ is the normalized mode profile, i.e., $\int_{-\infty}^{+\infty} |\psi_k(x)|^2 dx = 1$. One can see that irrespectively of the value of period $d$ the projection is maximal for eigenmodes from the center of the band, hence excitation efficiency of such modes is maximal. The example of the linear mode corresponding to maximal projection $I_k$ is depicted in Fig. 3(c), where we show only the vicinity of the left edge of the array. Such an eigenmode has well-defined peak in the left outermost pillar, that strongly overlaps with pump profile. Because different linear modes has close eigenvalues and comparable projections $I_k$, in the absence of nonlinearity localized pump will always excite their linear combinations, where the weight of the particular mode $k$ will be maximal (and spatial profile $\Psi$ will be close to $\psi_k$), when pump frequency will coincide with the eigenvalue $\varepsilon_k$.

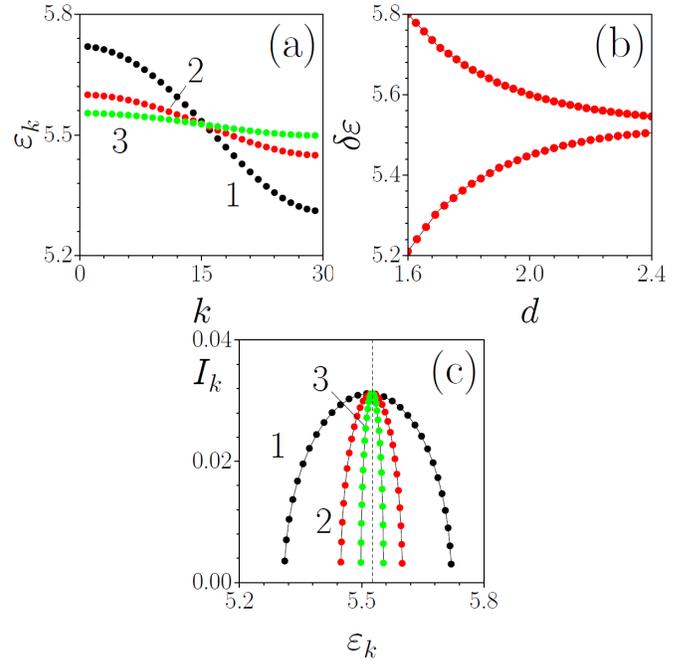

Figure 1. (a) Eigenvalues of the modes of the waveguide array with $d = 1.7$ (dots 1), $2.0$ (dots 2), and $2.3$ (dots 3). (b) Width of the first band of eigenvalues versus spacing $d$. (c) Projection of the localized pump on linear eigenmodes of the system from the first band at $h = 1$ and $d = 1.7$ (dots 1), $2.0$ (dots 2), and $2.3$ (dots 3).

When nonlinearity enters the scene, it substantially enriches the variety of available states and makes the formation of surface solitons possible. Looking for the steady-state solutions of the nonlinear version of Eq. (1) in the form $\Psi = \psi(x)e^{i\varepsilon t}$, one obtains the equation for the surface soliton profile

$$\frac{1}{2}\frac{d^2\psi}{dx^2} - \mathcal{R}(x)\psi - |\psi|^2\psi + i\gamma\psi - \mathcal{H}(x) - \varepsilon\psi = 0. \qquad (3)$$

To find its solutions, we applied Newton relaxation method. We monitor the peak amplitude $a_{\max} = \max|\psi|$ and the integral soliton width $w = U^2/\int_{-\infty}^{+\infty}|\psi|^4 dx$, where $U = \int_{-\infty}^{+\infty}|\psi|^2 dx$ is the soliton norm, as functions of the pump frequency detuning $\varepsilon$. Corresponding resonance curves $a_{\max}(\varepsilon)$ and $w(\varepsilon)$ are shown in Fig. 2, while examples of excited surface solitons are shown in Fig. 3. The key feature of $a_{\max}(\varepsilon)$ curves is bistability appearing at sufficiently large pump amplitudes $h$ due to nonlinearity-induced tilt of resonances, leading at large $h$ to shift of the position of resonance peak from the upper band [Fig. 1(a)] into forbidden energy gap. This means that Bragg reflection in periodic potential plays considerable role in localization of corresponding surface states. Notice that for selected strength of losses $\gamma$ and array parameters the width of the resonance curves $a_{\max}(\varepsilon)$ in Figs. 2(a) and 2(c) is comparable with the width of the entire upper band in the linear spectrum of the array, i.e. contributions from many modes are simultaneously present in soliton profile that is nevertheless clearly localized at the left edge of the array, where pump acts (see Fig. 3). Figure 2(c) illustrates transformation of the resonance curves with increasing spatial period $d$ of the array at fixed pump amplitude. Growth of $d$ leads to diminishing of coupling between pillars, it reduces currents from surface pillar (contributing to effective dissipation in it), that finally leads to notable growth of peak amplitude $a_{\max}$.

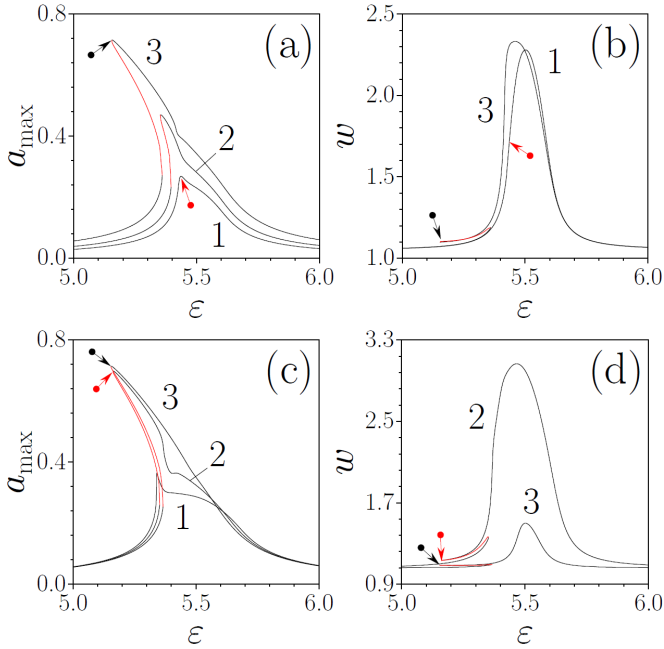
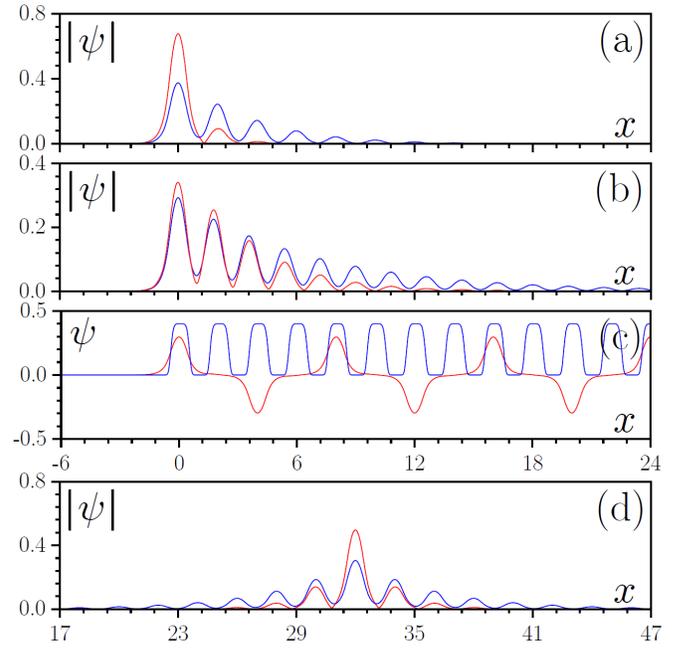

Figure 2. Peak amplitude (a) and width (b) versus detuning $\varepsilon$ at fixed $d=2.0$ and $h=0.015$ (curve 1), $0.020$ (curve 2), $0.029$ (curve 3). Peak amplitude (c) and width (d) versus detuning $\varepsilon$ at fixed $h=0.029$ and $d=1.8$ (curve 1), $1.9$ (curve 2), $2.2$ (curve 3). (d) Width of surface soliton versus detuning $\varepsilon$ at $h=0.029$ and $d=1.9$. Arrows indicate the points corresponding to resonance tips.

Figure 3. Surface soliton profiles in array with (a) $d=2.0$ at $\varepsilon=5.20$ (red curve) and $\varepsilon=5.50$ (blue curve), and array with (b) $d=1.8$ at $\varepsilon=5.34$ (red curve) and $\varepsilon=5.48$ (blue curve). In both cases $h=0.029$. (c) Profile of linear mode with $\varepsilon=5.52$ at $d=2.0$ (red curve) on which projection of surface soliton is maximal and $-\mathcal{R}(x)$ profile (blue curve). (d) Solitons in the center of array at $\varepsilon=5.36$ (red) and $\varepsilon=5.50$ (blue) for $h=0.029$.

The dependencies $w(\varepsilon)$ are entirely different: while point corresponding to maximal amplitude $a_{\max}$ [see arrows in Figs. 2(b) and 2(d)] corresponds to small soliton width and strong localization near the surface pillar, the maximum of width of the excited nonlinear state (indicating on its strong expansion into array) is achieved for detuning $\varepsilon$ close to the center of the upper band for all pump amplitudes $h$ [Fig. 2(b)]. Especially strong expansion into the depth of the array is possible for small periods $d$ [Fig. 2(d)]. Notice considerable reduction in the width of the $w(\varepsilon)$ dependence with increase of period $d$ that reflects shrinkage of the band [Fig.1(b)].

These soliton shape transformations are illustrated in Fig. 3(a) that shows surface soliton profiles for fixed period $d$ and two different values of frequency detuning, one of which correspond to the maximum of soliton amplitude ($\varepsilon=5.2$), while other corresponds to the maximum of the integral soliton width ($\varepsilon=5.5$). While in the former case the soliton is strongly localized near edge pillar, in the latter case it notably expands into array. It is an illuminating example of the possibility to control surface soliton amplitude and localization by tuning pump frequency. As mentioned above, the expansion of soliton into the depth of array can be much stronger for smaller spatial periods $d$ [Fig. 3(b)].

Linear stability analysis was performed by assuming the shapes of perturbed solitons in the form $\Psi=[\psi(x)+u(x)e^{\lambda t}+v^*(x)e^{\lambda^* t}]e^{i\varepsilon t}$ and substituting these expressions into Eq. (1), whose linearization yields the following linear eigenvalue problem:

$$i\lambda u = -\frac{1}{2}\frac{d^2 u}{dx^2} + \mathcal{R}(x)u + 2|\psi|^2 u + \psi^2 v - i\gamma u + \varepsilon u,$$
$$i\lambda v = +\frac{1}{2}\frac{d^2 v}{dx^2} - \mathcal{R}(x)v - 2|\psi|^2 v - \psi^{*2} u - i\gamma v - \varepsilon v,$$
(4)

for perturbation profiles $u,v \ll \psi$ and corresponding growth rates $\lambda$. Its solution has shown stability of the lower and upper branches of the resonance curves, even in the bistability regime, where resonances exhibit considerable tilts [see black branches in Figs. 2(a),(c)]. Middle branches are always unstable [red branches in Figs. 2(a),(c)].

Finally, we applied direct numerical integration of the Gross-Pitaevskii equation (1) to analyze dynamics of surface solitons formation and their stability beyond the perturbative approach. Figure 4 illustrates typical stable evolution of surface solitons from the upper branch of the resonance curve. These solitons were perturbed at $t=0$ by the broadband Gaussian noise with standard deviation $\sigma \sim 0.02$. Panel (a) corresponds to soliton with maximal amplitude (reached at $\varepsilon=5.2$), while panel (b) corresponds to state with maximal width ($\varepsilon=5.5$). In both cases, the perturbed soliton "cleans up" from the noise at typical times $\sim 5\gamma^{-1}$ due to dissipation that is consistent with results of the linear stability analysis. Unstable surface solitons from the middle branch (Fig.2, red curves) may jump into lower or upper branches depending on the level of the Gaussian noise $\sigma$: weak noise causes switching to the lower branch, while strong noise causes jump to the upper one.

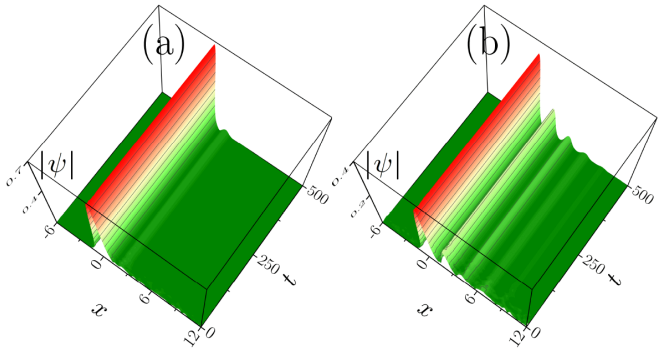

Figure 4. Stable evolution of perturbed polariton surface solitons from the upper branch with (a) $\varepsilon = 5.2$ and (b) $\varepsilon = 5.5$ at $h = 0.029$, $d = 2.0$.

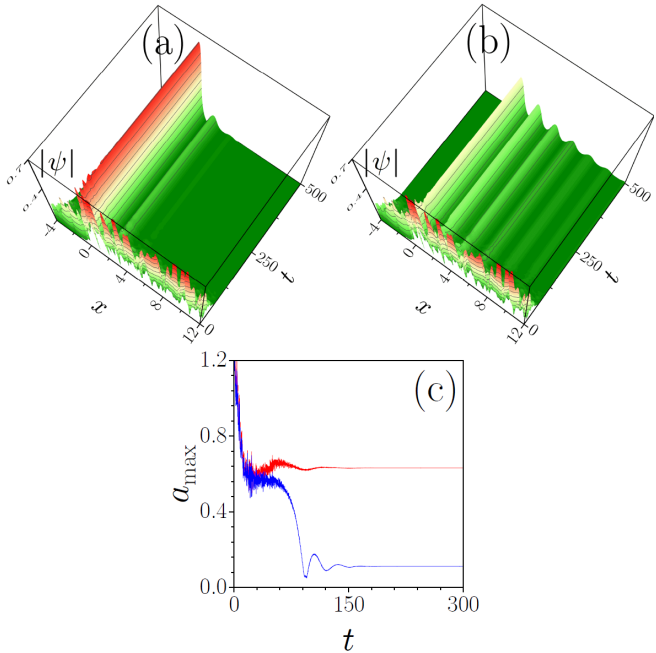

Figure 5. Generation of surface solitons from the upper (a) or lower (b) branch from random initial excitation at $\varepsilon = 5.25$, $h = 0.029$, $d = 2.0$ and (c) corresponding dependencies of peak amplitudes on time.

Similar phenomenon occurs upon spontaneous generation of surface solitons from seed Gaussian noise (Fig. 5). For selected in Fig. 5 noise realization, there exists threshold value of the standard deviation $\sigma_{\text{th}} = 0.56$, above which surface soliton from the upper branch is excited after relatively short time interval [Fig. 5(a)], while below it the soliton from the lower branch forms [Fig. 5(b)]. Figure 5(c) illustrates corresponding dynamics of peak amplitude $a_{\max}(t)$. Notice that $\sigma_{\text{th}}$ varies from one realization of noise to another typically within the range of 10%.

Notice that stable solitons can also form when pump is provided in the center of the array. In comparison with the surface ones, they are expectably symmetric, their width is about two times larger, while amplitude is slightly smaller for the same pump amplitude [Fig. 3(d)].

Summarizing, we uncovered salient features of the dissipative surface soliton formation in the exciton-polariton condensate in the one-dimensional array of microcavity pillars under the action of resonant pump localized in the edge resonator. The peak amplitude and width of resulting surface solitons could be effectively controlled by pump amplitude and frequency detuning. Importantly, surface solitons from the upper and lower branches are stable and could be formed from the seed Gaussian noise.